\documentclass[AMA,STIX1COL]{WileyNJD-v2}

\usepackage{lineno,hyperref}
\usepackage{graphicx}
\usepackage{amssymb}
\usepackage{amsmath}
\usepackage{listings}
\usepackage{cases}
\usepackage{amssymb}
\usepackage{array}
\usepackage{tabularx}
\usepackage{csquotes}
\usepackage{array,multirow}
\usepackage{hhline}
\usepackage{epsfig}
\usepackage{soul}
\usepackage{color}
\DeclareGraphicsExtensions{.pdf,.png,.jpg}
\usepackage{bm}
\usepackage{cleveref}
\usepackage{nameref}
\usepackage{amsmath}
\usepackage{algorithm}
\usepackage{tabularx}
\usepackage{float}
\usepackage{booktabs}

\usepackage{svg}

\definecolor{highlightgray}{gray}{0.9}

\graphicspath {{images/}}
\renewcommand\thesubsection{\Alph{subsection}}

\articletype{Research Article}%

\received{<day> <Month>, <year>}
\revised{<day> <Month>, <year>}
\accepted{<day> <Month>, <year>}

\begin{document}

\title{Experimental Assessment of Containers Running on Top of Virtual Machines}

\author[1]{Hossein Aqasizade}

\author[1,2]{Ehsan Ataie}

\author[1,2]{Mostafa Bastam}

\authormark{HOSSEIN AQASIZADE \textsc{et al}}

\address[1]{\orgdiv{Department of Computer Engineering}, \orgname{University of Mazandaran}, \orgaddress{\state{Babolsar}, \country{Iran}}}

\address[2]{\orgdiv{Distributed Computing Systems Research Group}, \orgname{University of Mazandaran}, \orgaddress{\state{Babolsar}, \country{Iran}}}

\corres{*Ehsan Ataie. Department of Computer Engineering, University of Mazandaran, Babolsar, Iran. \email{ataie@umz.ac.ir}}

	\abstract[Summary]{Over the past two decades, the cloud computing paradigm has gradually attracted more popularity due to its efficient resource usage and simple service access model. Virtualization technology is the fundamental element of cloud computing that brings several benefits to cloud users and providers, such as workload isolation, energy efficiency, server consolidation, and cost reduction. 
	This paper examines the combination of operating system-level virtualization (containers) and hardware-level virtualization (virtual machines). To this end, the performance of containers running on top of virtual machines is experimentally compared with standalone virtual machines and containers based on different hardware resources, including the processor, main memory, disk, and network in a real testbed by running the most commonly used benchmarks. Paravirtualization and full virtualization as well as type 1 and type 2 hypervisors are covered in this study. In addition, three prevalent containerization platforms are examined.}


\maketitle

\section{Introduction}\label{sec1}

Cloud computing is constantly evolving to meet the growing demand for computing resources. This paradigm is fundamentally built upon virtualization technologies. 
Cloud service providers exploit virtualization to enhance resource efficiency and management, cost savings, reliability, portability, energy efficiency, and reduction of data center expenses \textcolor{blue}{\cite{randal2020ideal,verma2021auto,masdari2020efficient}}. Thanks to this popularity, virtualization has been an interesting topic for researchers for many years.

Broadly speaking, there are two main virtualization methods used in cloud systems: hardware-level virtualization and operating system (OS)-level virtualization \textcolor{blue}{\cite{felter2015updated}}. 
Hardware-level virtualization employs virtual machines (VMs) to create virtualized instances of the physical environment. In general, there are two types of virtual machine monitors (VMMs), also known as hypervisors: \textit{type 1} and \textit{type 2}. Type 1 hypervisors are installed directly on hardware and provide a form of operating system or kernel, while type 2 ones are installed as an application on top of an existing OS. 
In contrast to hardware-level virtualization, containers in the operating system-level model share libraries, drivers, or OS-dependent binaries with the host kernel. In other words, OS-level virtualization refers to the feature that enables different virtual instances of the user space within the kernel space.

Although there are several virtualization techniques, full virtualization and paravirtualization (PV) are the most well-known.
Full virtualization provides a complete simulation of the underlying hardware, allowing guest operating systems to operate as if they have direct access to the machine’s physical hardware via the hypervisor, while paravirtualization interacts more efficiently with the hypervisor through a dedicated API, necessitating modifications to the guest OS for compatibility.
The advantages of full virtualization include portability, better compatibility, and support for guest operating systems without modification.
The advantages of paravirtualization include improved performance, better resource management, enhanced scalability, and compatibility with older hardware.

Various VMMs are prevalent in today's world, with KVM \textcolor{blue}{\cite{kvm}} and Xen \textcolor{blue}{\cite{xen}} being among the most well-known. 
While KVM primarily supports full virtualization technique, Xen has introduced a hybrid technique known as paravirtual hardware virtual machine (PVHVM), which combines elements of paravirtualization and hardware-assisted virtualization. The primary goal of PVHVM is to enhance the performance of fully virtualized HVM guests by incorporating optimized paravirtualization drivers. Specifically, PVHVM uses paravirtualized drivers within HVM environments to bypass the overhead associated with emulating disk and network operations, thus improving I/O efficiency.

A container as an alternative to a VM is a standardized software unit that bundles the code and all its dependencies so that a program can be executed quickly, reliably and with high portability on a container engine using operating system kernel services\textcolor{blue}{\cite{hyder2023toward, ouyang2023container, zhangoptimal}}. 
On the other hand, the main disadvantage of container-based virtualization technologies is poor isolation \textcolor{blue}{\cite{mavridis2023orchestrated}}. 
Some of the best-known containerization platforms are LXC \textcolor{blue}{\cite{lxc}}, Docker \textcolor{blue}{\cite{docker}}, and Podman \textcolor{blue}{\cite{podman}}.
LXC, representing an earlier generation of containers, combines the Linux kernel's cgroups and support for isolated namespaces to provide an isolated environment for applications. 
Docker is a portable, lightweight and widely used container-based technology that automates the deployment of applications in containers and focuses primarily on the application containerization with an image-centric management model \textcolor{blue}{\cite{pratap2021formal, tang2022container,he2023real}}.
Podman is a container engine that operates without requiring system-level access and has been developed by RedHat \textcolor{blue}{\cite{redhat}}. It has been introduced as an alternative to the widely-known Docker platform. 

\begin{figure}[h]
	\centering
	\includegraphics[width=.8\textwidth]{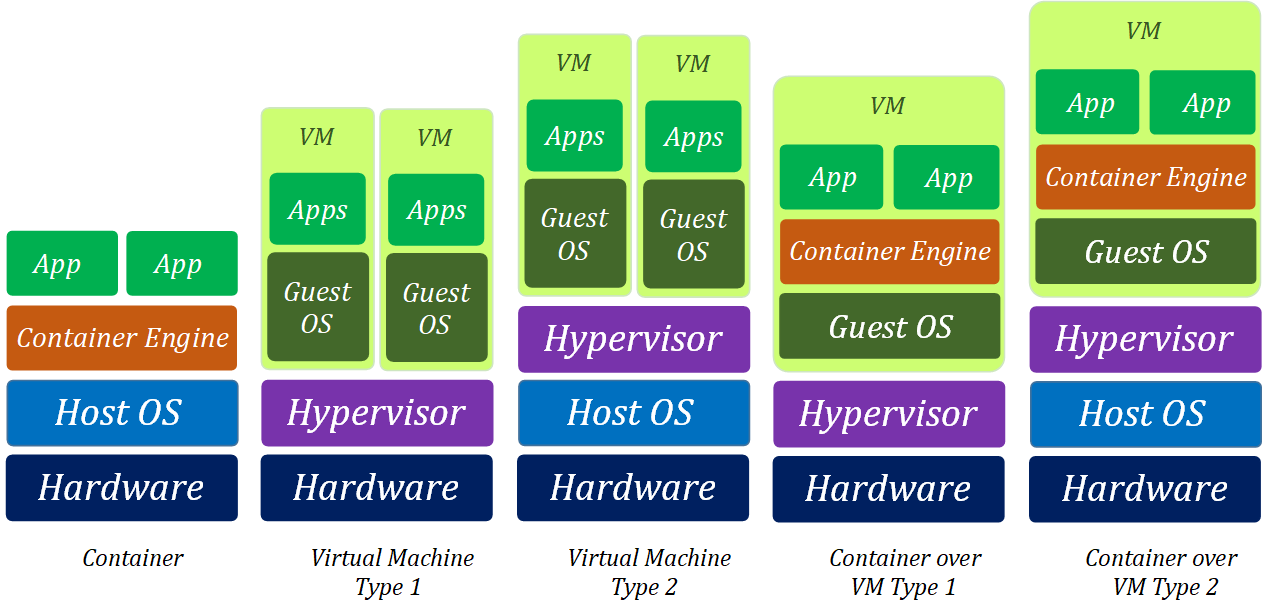}
	\captionsetup{justification=centering}
	\caption{General architecture of containers, virtual machines, and their combination}\label{fig1}
\end{figure}

This paper examines the performance of VMs, containers, and containers on top of VMs from various aspects. To this end, Xen and KVM are used as type 1 and type 2 hypervisors, respectively. In addition, LXC, Docker, and Podman are exploited as container technologies. By combining these VM and container platforms, several configurations for the execution of containers on VMs are defined and tested with known and common benchmarking tools.
Various metrics are used to assess the performance of the configurations when using the processor, memory, disk, and network. To the best of our knowledge, this study is the most comprehensive among the few existing works that have investigated the performance of containers running on VMs versus standalone containers and standalone VMs, considering different types of hypervisors and containerization platforms. The result of this study provides cloud data center researchers and developers with a comprehensive overview that enables them to select the best possible configuration for their operating environment based on application requirements. Figure \textcolor{blue}{\ref{fig1}} shows the architecture of containers, virtual machines and their combination.

The rest of this paper is organized as follows: Section 2 gives an overview of related research in the field. Section 3 discusses the software, hardware and benchmarks used and presents the configuration scenarios investigated in this research. Section 4 compares different configurations and presents the experimental evaluation results. Finally, Section 5 concludes the paper with some directions for future work.

%%%%%%%%%%%%%%%%%%%%%%%%%%%%%%%%%%%%%%%%%%%%%%%%%%%%%
%      Related Work
%%%%%%%%%%%%%%%%%%%%%%%%%%%%%%%%%%%%%%%%%%%%%%%%%%%%%

\section{Related Work}\label{sec2}

The literature review in this article is divided into three categories that highlight the importance of virtualization technologies for researchers. The first category analyzes studies that focus on virtual machines launched on different hypervisor platforms. The second category examines studies that focus on container-based virtualization. Finally, the third category comprises studies that compare virtual machines and containers or examine containers on virtual machines. The literature review is intended to be brief while emphasizing the research relevant to this article.

\subsection{2.1. Virtual Machines}\label{sub2sec1}

{\DH}or{\dj}evi{\'c} et al. \textcolor{blue}{\cite{dhordjevic2020vmware}} have investigated the Hyper-V and ESXi hypervisors. Hyper-V outperformed ESXi in file system benchmarks when running under Windows Server 2016 as a guest OS, while ESXi showed slightly lower performance when running three virtual machines simultaneously. The difference between ESXi and Hyper-V was significant, with ESXi performing more than 50\% better, especially when the guest OS was CentOS.
{\DH}or{\dj}evi{\'c} et al. \textcolor{blue}{\cite{djordjevic2021performance}} have compared the performance of a native machine with the Oracle VirtualBox hypervisor using the Filebench software. The study was conducted with one, two, and three VMs running on VirtualBox and shows that the native machine performs much better in file system related experiments.

Elsayed et al. \textcolor{blue}{\cite{elsayed2013performance}} have evaluated three well-known hypervisors, VMware ESXi 5, Microsoft Hyper-V 2008 R2, and Citrix Xen Server 6.0.2. The study revealed that VMware ESXi 5 achieved significant and acceptable results in SQL performance in terms of processor and disk, followed by Hyper-V 2008 R2 and Citrix Xen Server 6.0.2. However, at low workloads, Citrix Xen 6.0.2 outperformed Hyper-V 2008 R2 and VMware ESXi 5. As resource consumption increased, the performance of Citrix Xen Server 6.0.2 suffered from disk saturation.
Reddy et al. \textcolor{blue}{\cite{reddy2014performance}} have investigated the performance evaluation of VMWare ESXi Server, XenServer, and KVM in private cloud environments. The results demonstrated that XenServer and ESXi Server performed comparably well, coming very close to the performance of native machines without significant virtualization overhead. In the processor tests, ESXi outperformed XenServer, while XenServer performed better in the memory and I/O tests.
{\DH}or{\dj}evic et al. \textcolor{blue}{\cite{dhordjevic2020performance}} have compared four well-known hypervisors, namely KVM, ESXi, Xen, and Hyper-V, from a file system perspective. The study concluded that the performance of the native machine was superior in all file system benchmarks, which was primarily due to the fact that the native machines used cache, which was not the case with KVM.

\subsection{2.2. Containers}\label{sub2sec2}
Moravcik et al. \textcolor{blue}{\cite{moravcik2020comparison}} have presented a comparison between two popular container platforms, LXC and Docker. From the researchers’ point of view and based on their investigations, both platforms demonstrated almost equivalent performance, although Docker had slightly higher request round-trip latency compared to LXC. Nevertheless, each technology has its own advantages for certain use cases.
Gantikow et al. \textcolor{blue}{\cite{gantikow2020rootless}} have found that Podman is particularly suitable for HPC environments. According to the research, Podman offers beneficial features such as non-root containers for HPC, user-only container image creation, pod support, and stronger isolation. 
Bachiega et al. \textcolor{blue}{\cite{bachiega2020performance}} have investigated the performance of data sharing with Docker Volume and NFS. The study showed that Docker has a simpler and more dynamic configuration compared to NFS. However, NFS demonstrates better asynchronous performance in all tests, which includes read and write operations on HDD and SSD disks.

Saxena et al. \textcolor{blue}{\cite{saxena2021analysis}} have focused on investigating Docker containers and their performance and importance in bare metal environments was emphasized. The measurements indicated that containers perform better when running on native machines than in situations where they are running on top of a VM.
{\DJ}or{\dj}evi{\'c} et al. \textcolor{blue}{\cite{djordjevic2022performance}} have contrasted two container-based virtualizations, Podman and Docker.
Using various workload files from the Filebench benchmark, the authors demonstrated that the performance difference between Docker and Podman is essentially nonexistent.
The results also showed that the amount of overhead from Docker and Podman has a negligible impact on performance.
{Peri{\'c} et al. \textcolor{blue}{\cite{peric2022analysis}} have investigated the minimum system requirements for running Android OS inside Linux containers. They identified a minimal configuration of Android running inside a Linux container as a suitable solution for resource-constrained embedded systems. The authors also found that developing Android apps inside LXC may be an effective solution for developers without access to high-performance computers.

\subsection{2.3. Containers and Virtual Machines}\label{sub2sec3}

Zhang  et al. \textcolor{blue}{\cite{zhang2018comparative}} have conducted an in-depth empirical study to compare virtual machines and Docker containers in a Spark environment. The authors came to three interesting conclusions. Firstly, Docker containers are easier for system administrators to implement and start compared to virtual machines. Secondly, Docker containers have much better scalability than virtual machines for different big data sets. Finally, Docker containers achieve higher efficiency in CPU and disk usage for the same size of workload.
Sharma  et al. \textcolor{blue}{\cite{sharma2016containers}} have compared virtual machines and containers, leading to the conclusion that unlike virtual machines with hard resource constraints, containers also have soft resource constraints, which can be beneficial in overcommitment scenarios. The study showed that Docker uses a copy-on-write file system and version control, which facilitates continuous deployment and better integration. According to the authors, while lightweight virtual machines such as Clear Linux are popular for container-like features, approaches that combine virtualization techniques seem promising due to the separation characteristics of virtual machines.
 
Ramalho et al. \textcolor{blue}{\cite{ramalho2016virtualization}} have compared the performance of container-based and hypervisor-based approaches in IoT and edge network devices. Artificial benchmarks were used on a Cubieboard2 SoC platform. The results demonstrated a significant overhead for the hypervisor-based solution, which was not easily reducible. However, the Docker platform showed promising results and reduced the overhead significantly.
Chae et al. \textcolor{blue}{\cite{chae2019performance}} have compared Docker containers and KVM virtual machines on a single machine. They found that KVM is significantly less efficient and requires 3.6 to 4.6 times more memory for each VM due to the complete instantiation of the operating system. Docker proved to be more resource efficient, especially in CPU, HDD, and RAM usage. KVM also suffered from slow VM creation, which affected scalability.
 
Mavridis et al. \textcolor{blue}{\cite{mavridis2019combining}} have researched the combination of containers and VMs in cloud computing. The performance evaluations have shown that container technologies running on VMs have minimal overheads in memory usage and CPU. Data volume options were also highlighted, and the study provided insights into the power consumption and energy overhead of running containers on VMs. Overall, the results demonstrated the efficiency and effectiveness of using containers within VMs for cloud computing environments.
Mondesire et al. \textcolor{blue}{\cite{mondesire2019combining}} have extended traditional HPC to a cloud-based service that enables simultaneous interactive simulations and compute-intensive tasks. Four HPC load balancing techniques were investigated, the choice of which depends on cluster resources, software jobs and type. It was found that containerization, especially Docker, is preferred over virtual machines for game-based simulations. Docker demonstrated superior performance by minimizing overhead compared to virtual machines. The study highlights the potential of Docker to reduce overhead and increase performance, confirming its suitability for such computational workloads.

Lingayat et al. \textcolor{blue}{\cite{lingayat2018performance}} have investigated the performance impact of deploying Docker containers on different environments, including bare metal and VMs. The study found a significant performance improvement of around 50\% when running Docker containers on bare metal as opposed to VMs. This improvement is attributed to the additional layers introduced by the emulation-based virtual machine architecture. The study concludes that deployment on bare metal machines is recommended to take full advantage of Docker containers.
Mavridis et al. \textcolor{blue}{\cite{mavridis2017performance}} have evaluated the performance of containers running on VMs in cloud computing. Docker and the KVM hypervisor were used in experiments on various operating systems, where a performance overhead due to the additional virtualization layer was observed. The study emphasizes that while VMs enhance security and management, they also come with performance penalties.
The decision between VMs and bare metal depends on the specific application requirements, with VMs recommended for security and management and bare metal for faster systems with lower overhead.

To gain a deeper understanding, Table \textcolor{blue}{\ref{tab1}} compares our work and other work reviewed in Section 2.3 based on the category, platforms used, and resources evaluated.

\begin{table}[h]
	\centering
	\caption{Our work compared to the related work discussed in Section 2.3}\label{tab1}
	\begin{center}
		\begin{tabular}{@{}llll>{\centering\arraybackslash}p{0.7cm}>{\centering\arraybackslash}p{1.2cm}>{\centering\arraybackslash}p{0.7cm}>{\centering\arraybackslash}p{1.4cm}@{}}
			\toprule
			\textbf{Related Work} & \textbf{Category} & \textbf{VMM} & \textbf{Container Platform} & \multicolumn{4}{c}{\textbf{Resource}} \\
			\cmidrule(lr){5-8}
			& & & & \textbf{CPU} & \textbf{Memory} & \textbf{Disk} & \textbf{Network} \\
			\midrule
			{Zhang et al.} \textcolor{blue}{\cite{zhang2018comparative}} & VM vs. container & KVM & Docker & \checkmark & \checkmark  \\
			{Sharma et al.} \textcolor{blue}{\cite{sharma2016containers}} & VM vs. container & KVM & LXC & \checkmark & \checkmark & \checkmark & \checkmark\\
			{Ramalho et al.} \textcolor{blue}{\cite{ramalho2016virtualization}} & VM vs. container & KVM & Docker & \checkmark & \checkmark & \checkmark & \checkmark\\
			{Chae et al.} \textcolor{blue}{\cite{chae2019performance}} &  VM vs. container & KVM & Docker & \checkmark & \checkmark & \checkmark & \\
			{Mavridis et al.} \textcolor{blue}{\cite{mavridis2019combining}} & container on VM & KVM, Xen, Hyper-V & Docker & \checkmark & \checkmark & \checkmark & \checkmark\\
			{Mondesire et al.} \textcolor{blue}{\cite{mondesire2019combining}} & container on VM & VMware Workstation & Docker & \checkmark & \checkmark & & \checkmark\\
			{Lingayat et al.} \textcolor{blue}{\cite{lingayat2018performance}} & container on VM & KVM & Docker & & \checkmark & & \\
			{Mavridis et al.} \textcolor{blue}{\cite{mavridis2017performance}} & container on VM & KVM & Docker & \checkmark & \checkmark & \checkmark & \checkmark \\
			Our work & container on VM & KVM, Xen & Docker, LXC, Podman & \checkmark & \checkmark & \checkmark & \checkmark\\			
			\bottomrule
		\end{tabular}
	\end{center}
\end{table}

%%%%%%%%%%%%%%%%%%%%%%%%%%%%%%%%%%%%%%%%%%%%%%%%%%%%%
%      Research Methodology
%%%%%%%%%%%%%%%%%%%%%%%%%%%%%%%%%%%%%%%%%%%%%%%%%%%%%

\section{Research Methodology}\label{sec3}

In this section you will find general information on software and hardware, benchmarks, configuration modes, and execution methods used in this research.

\subsection{3.1. Software and Hardware Configuration}\label{sub3sec1}

All measurements are carried out on a computer with an Intel CoreTM i7-7700 processor @ 4.2 GHz CPU, an ASUS PRIME H270-PLUS motherboard with an Intel Xeon E3-1200 chipset, 8 GB RAM, a 1 TB WD10EZEX-75W hard disk, and a Realtek RTL8111/8186/8411 network card. The operating system used for all containers and virtual machines is Ubuntu Server 20.04~\textcolor{blue}{\cite{ubuntu}}.

The Xen server is set up and run as a type 1 hypervisor using XCP-ng version 8.2.1. XCP-ng is a Linux distribution of the Xen project that relies on the Xen hypervisor and the Xen API (XAPI) project and is an open source alternative to XenServer (Citrix hypervisor). XCP-ng aims to provide an easy way to deploy a virtualized infrastructure using Xen as a hypervisor. In this study, the Xen Orchestra application is run on Docker to manage virtual machines on the Xen server. Xen Orchestra is a web-based application designed for XenServer management, backup, and cloud initiation. It offers a unified interface that works across devices to control repositories, hosts, and virtual machines, enabling efficient management of the XenServer infrastructure.
The Xen server is run with PVHVM mode, which combines hardware-assisted virtualization with paravirtualization drivers for network and disk access. 

The second hypervisor examined in this study is KVM version 6.2, running as type 2 in HVM mode. A bridge mode is used for KVM virtual machines’ networking. The use of a Linux bridge in KVM enables the virtual machine to access networks and external services outside the virtual environment. With bridged networking, a virtual machine is assigned a dedicated network card that provides a communication path to external networks. Docker version 20.10.18, Podman version 3.4.4, and LXC version 5.5 are used to create containers. To ensure consistent measurement, all virtual machines are assigned the maximum available resources.

\subsection{3.2. Benchmarks}\label{sub3sec2}

In this study, the performance of the four most important hardware components - processor, memory, hard disk, and network - is examined. Benchmarking is a common practice for evaluating and  analyzing performance using established software packages known as benchmarks. This section describes the benchmarks used in the study to evaluate the performance of the various hardware components and systems. %The following well-known benchmarks are used in this study:

Compress-7zip~\textcolor{blue}{\cite{7zip}} version 1.10.0 is used to measure processor performance. It evaluates the processor performance by compressing and decompressing a 128 MB file with a 32 MB dictionary size. The benchmark specifies the number of million instructions per second (MIPS) for compression and decompression operations. A higher MIPS value indicates better performance.
STREAM~\textcolor{blue}{\cite{stream}} version 1.3.3 is a simple synthetic benchmark that measures sustained memory bandwidth in megabytes per second (MB/s). The benchmark requires each array size to be at least four times the size of all last-level cache memories used during execution or one million elements (whichever is larger). STREAM uses four types of operations: Copy, Scale, Add, and Triad. Table \textcolor{blue}{\ref{tab2}} shows the STREAM operations formulas.

\begin{table}[h]
	\centering 
	\caption{STREAM operations}\label{tab2}
	\begin{tabular}{lll}
		\toprule
		\textbf{Operation Type} & \textbf{Formula} \\
		\midrule
		Copy    & a(i) = b(i) \\
		Scale    & a(i) = q * b(i) \\
		Addition    & a(i) = b(i) + c(i)  \\
		Triple    & a(i) = b(i) + q * c(i) \\
		\bottomrule
	\end{tabular}
\end{table}
IOzone~\textcolor{blue}{\cite{iozone}} is a file system benchmarking tool that evaluates disk performance by generating various file operations. This benchmark assesses the input/output (I/O) performance by reading and writing to the disk. For this purpose, a 512 MB file with a 2 MB record size is examined.
Netperf~\textcolor{blue}{\cite{netperf}}  is a network benchmarking tool that evaluates network performance from various aspects and measures the network bandwidth between two hosts. It is also suitable for determining the latency between two hosts. Netperf offers various tests, and in this work, TCP STREAM is used to measure the bandwidth of the TCP protocol. The way it works is that a certain amount of data is transferred from the system running Netperf to the system running Netserver. In addition, TCP Request-Response and UDP Request-Response tests are conducted in this study. A TCP Request-Response test can be considered as a user space ping. The transaction rate is the number of fully exchanged transactions divided by the time it takes to complete these transactions. The UPD Request-Response test is very similar to the TCP Request-Response test, except that UDP is used instead of TCP. UDP does not provide the ability to retransmit lost UDP datagrams, nor does Netperf add anything. This means that if a request or response is lost, the exchange of requests and responses is stopped from that point until the end of the test period.

Furthermore, Sysbench's \textcolor{blue}{\cite{sysbench}} Online Transaction Processing (OLTP) test is used to evaluate the performance of MySQL~\textcolor{blue}{\cite{mysql}}. This test performs operations such as inserting, updating, deleting or querying data in a database, with the most important output being transactions per second (TPS). The OLTP test focuses specifically on key lookup queries. It involves creating a user, granting access, generating tables, and assessing performance with different thread counts.

\subsection{3.3 Configuration Scenarios}\label{sub3sec3}

For the purpose of performance evaluation, twelve different configuration scenarios are employed to assess computational performance. Table \textcolor{blue}{\ref{tab3}} illustrates these configuration scenarios.
While the first six scenarios, including native machine, standalone VMs, and standalone containers, mainly serve as the baseline for performance evaluation, the second six scenarios are the main scenarios that provide different configurations for running containers on VMs.

\begin{table}[h]
	\centering
	\caption{Configuration scenarios examined in this study}\label{tab3}
	\begin{tabular}{lll}
		\toprule
		\textbf{No.} & \textbf{Scenario} & \textbf{Abbreviation} \\
		\midrule
		1    & Native Machine &  \textit{NTM}\\
		2    & KVM-based VM  & \textit{KVM} \\
		3    & Xen-based VM  & \textit{XVM} \\
		4    & Docker container & \textit{DKC} \\
		5    & Podman container & \textit{PDC}  \\
		6    & LXC container & \textit{LXC}  \\
		7    & Docker container on KVM VM & \textit{DoK}  \\
		8    & Podman container on KVM VM & \textit{PoK}  \\
		9    & LXC container on KVM VM & \textit{LoK} \\
		10   & Docker container on Xen VM & \textit{DoX}  \\
		11   & Podman container on Xen VM & \textit{PoX} \\
		12   & LXC container on Xen VM & \textit{LoX} \\	
		\bottomrule
	\end{tabular}
\end{table}

%%%%%%%%%%%%%%%%%%%%%%%%%%%%%%%%%%%%%%%%%%%%%%%%%%%%%
%      Results
%%%%%%%%%%%%%%%%%%%%%%%%%%%%%%%%%%%%%%%%%%%%%%%%%%%%%

\section{Results}\label{sec4}

In this section, the performance of the four main resources, i.e. processor, memory, disk, and network, as well as the performance of MySQL, are measured and analyzed when different scenarios are set up and run. The winning configurations for each resource are then summarized.

\subsection{4.1. Processor}\label{sub4sec1}
Processor performance is evaluated with the Compress-7zip benchmark for the native machine, virtual machines, containers and containers on virtual machines.
If we look at compression and decompression together, as shown in Figure \textcolor{blue}{\ref{fig2}}, the native machine comes first. This is followed by Podman, Docker and LXC containers.
After that, scenarios with containers over VMs show competitive performance with traditional VM setups, with DoK, LoK and PoX being the best scenarios respectively.
This suggests that Docker containers on KVM might be the best choice among the other container-over-VM scenarios if CPU performance is the main concern, though other configurations show almost equivalent performance.

\begin{figure}[!ht]
	\centering
	\begin{minipage}[c]{.67\linewidth}
		\includegraphics[width=\linewidth]{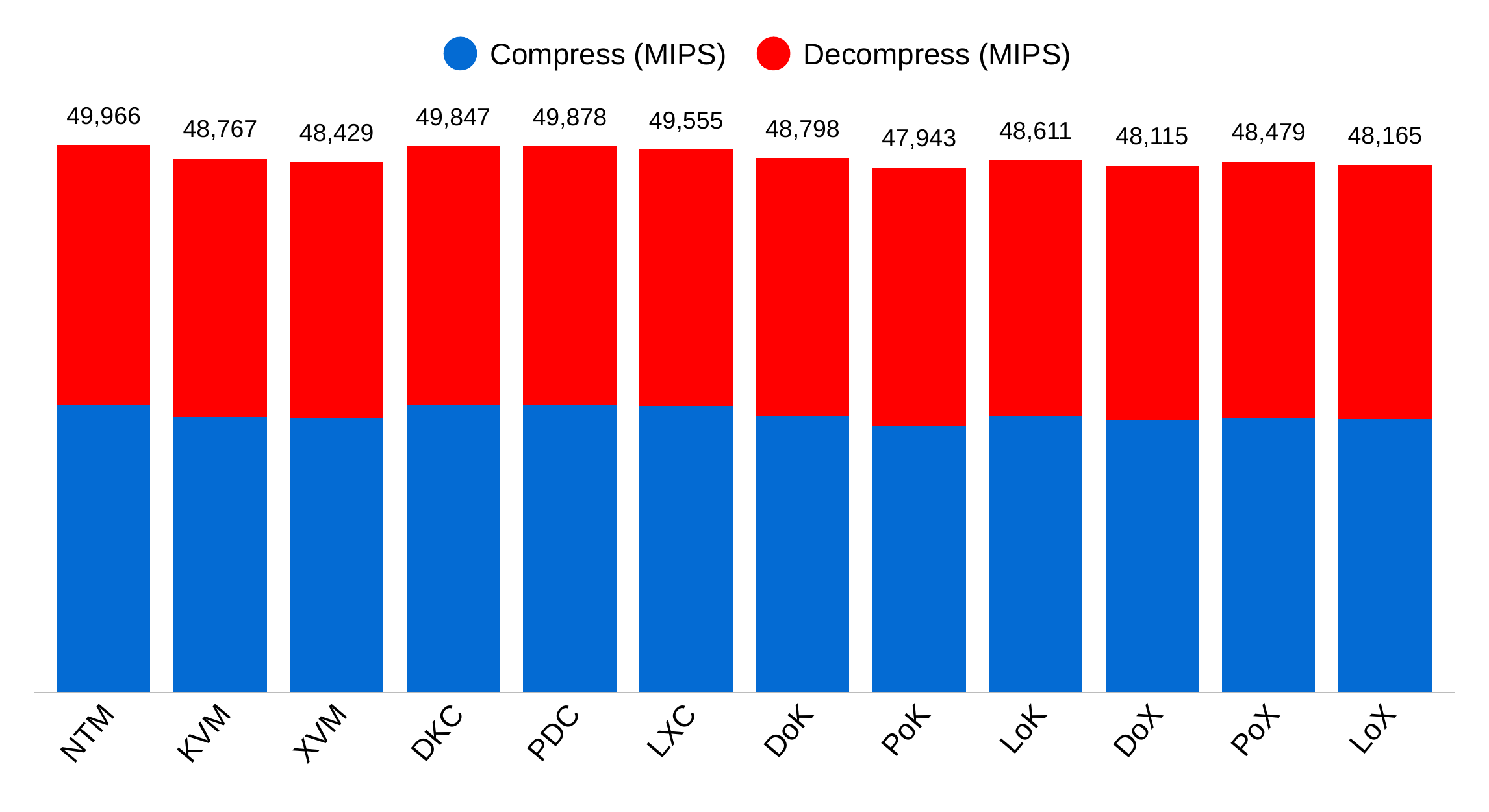}
	\end{minipage}%
	\begin{minipage}[c]{.33\linewidth}
		\captionsetup{justification=raggedright, singlelinecheck=false}
		\makebox[0pt][l]{\parbox{\linewidth}{%
				\caption{Compress-7zip test results}\label{fig2}
		}}
	\end{minipage}
\end{figure}

\subsection{4.2. Memory}\label{sub4sec2}
Memory performance is measured using the STREAM benchmark, which comprises four operations: Copy, Scale, Triad and Add. Figure \textcolor{blue}{\ref{fig3}} represents the comparison of memory performance between different configurations. At first glance, all configuration scenarios show similar performance due to their respective virtualization layers. However, a closer look reveals that, similar to the processor performance, the native machine has the highest memory performance in all configurations. Podman,
Docker, and LXC containers follow with little difference, and the remaining scenarios rank closely behind them. 

Among the container-over-VM configurations, scenarios with containers on Xen-based VMs, such as LoX and PoX, followed by DoX, demonstrate notable performance. These configurations exhibit competitive results in the STREAM benchmarks, suggesting that containers, whether LXC, Podman or Docker, can work seamlessly within the isolation framework provided by Xen virtualization.
On the other hand, LoK and PoK perform the worst not only among the other container-over-VM scenarios, but also among all other scenarios. This suggests that the KVM virtual machine is not a suitable candidate for hosting containers, especially LXC and Podman, when memory performance is critical.

\begin{figure}[!ht]
	\centering
	\begin{minipage}[c]{.3\linewidth}
	\captionsetup{justification=raggedright, singlelinecheck=false}
	\makebox[0pt][l]{\parbox{\linewidth}{%
			\caption{STREAM test results}\label{fig3}
	}}
	\end{minipage}%
	\begin{minipage}[c]{.7\linewidth}
		\includegraphics[width=\linewidth]{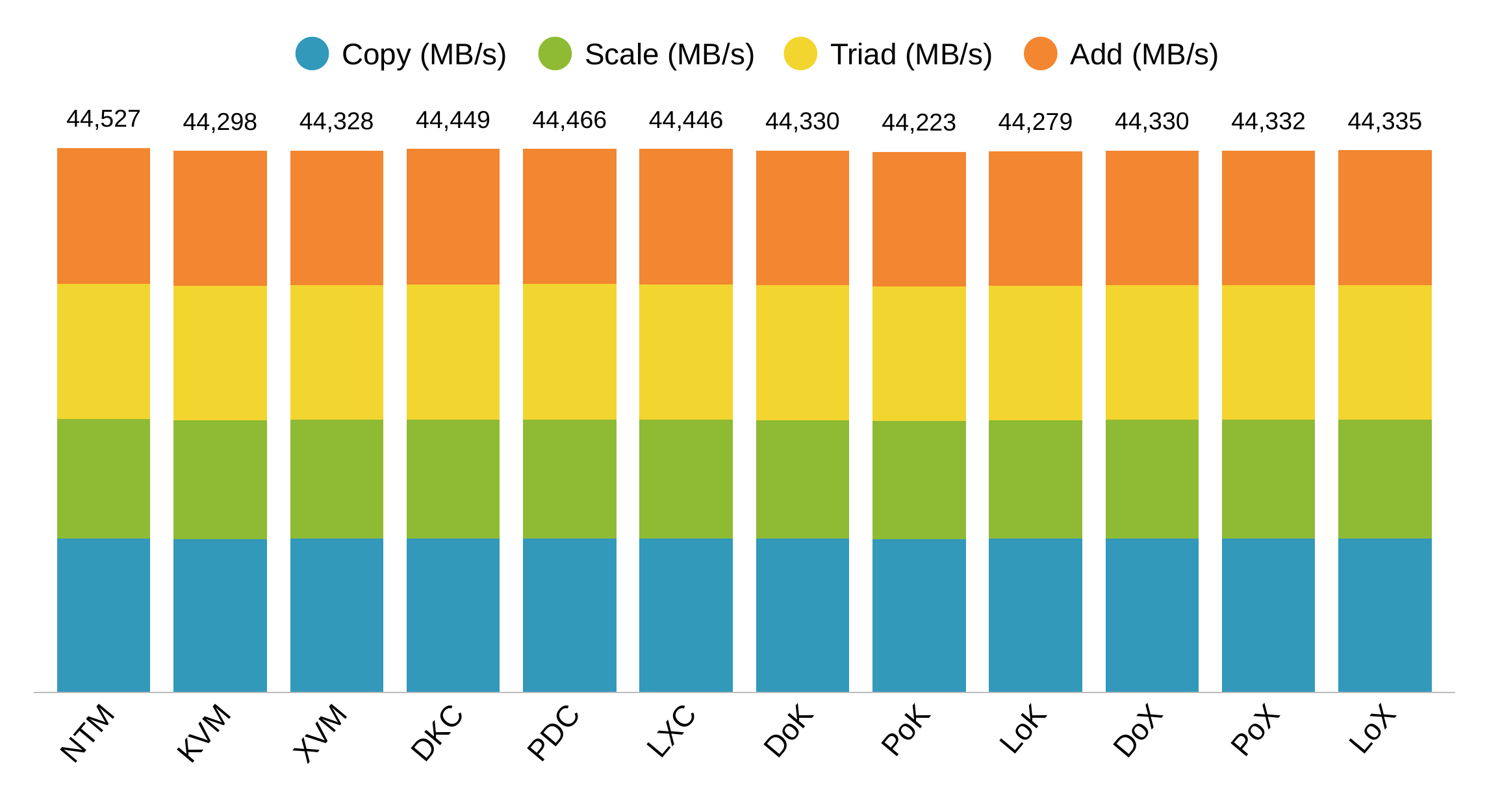} 
	\end{minipage}

\end{figure}

\subsection{4.3. Disk}\label{sub4sec3}

The hard disk performance is evaluated with the IOzone benchmark. The benchmark is run with a record size of 2 MB and a file size of 512 MB, and measures the read and write performance of the filesystem. Figures \textcolor{blue}{\ref{fig4}} and \textcolor{blue}{\ref{fig5}} show the results of the read and write tests for the hard disk. In contrast to the previous resources, the native machine does not show the highest performance for the read operations on the hard disk. Instead, the Xen-based VM achieves the highest performance. This is followed by container configurations on Xen-based VMs, with slight differences. The native machine comes next, followed by other configurations at a similar level.
The outperformance of Xen-based VM is primarily due to the use of the PVHVM technique. In PVHVM, paravirtualized drivers are used for the disk I/O, which results in better performance compared to the native machine, KVM-based VM, and containers. 

For write operations on the hard disk, the Xen-based VM shows the highest performance, similar to the read operations. The LoX and DoX configurations follow with a slight difference, followed by PoX.
The DoK and native machine come next. %After that, all configurations reach almost the same level.
\begin{figure}[h]
	\centering
	\begin{minipage}[c]{.7\linewidth}
		\includegraphics[width=\linewidth]{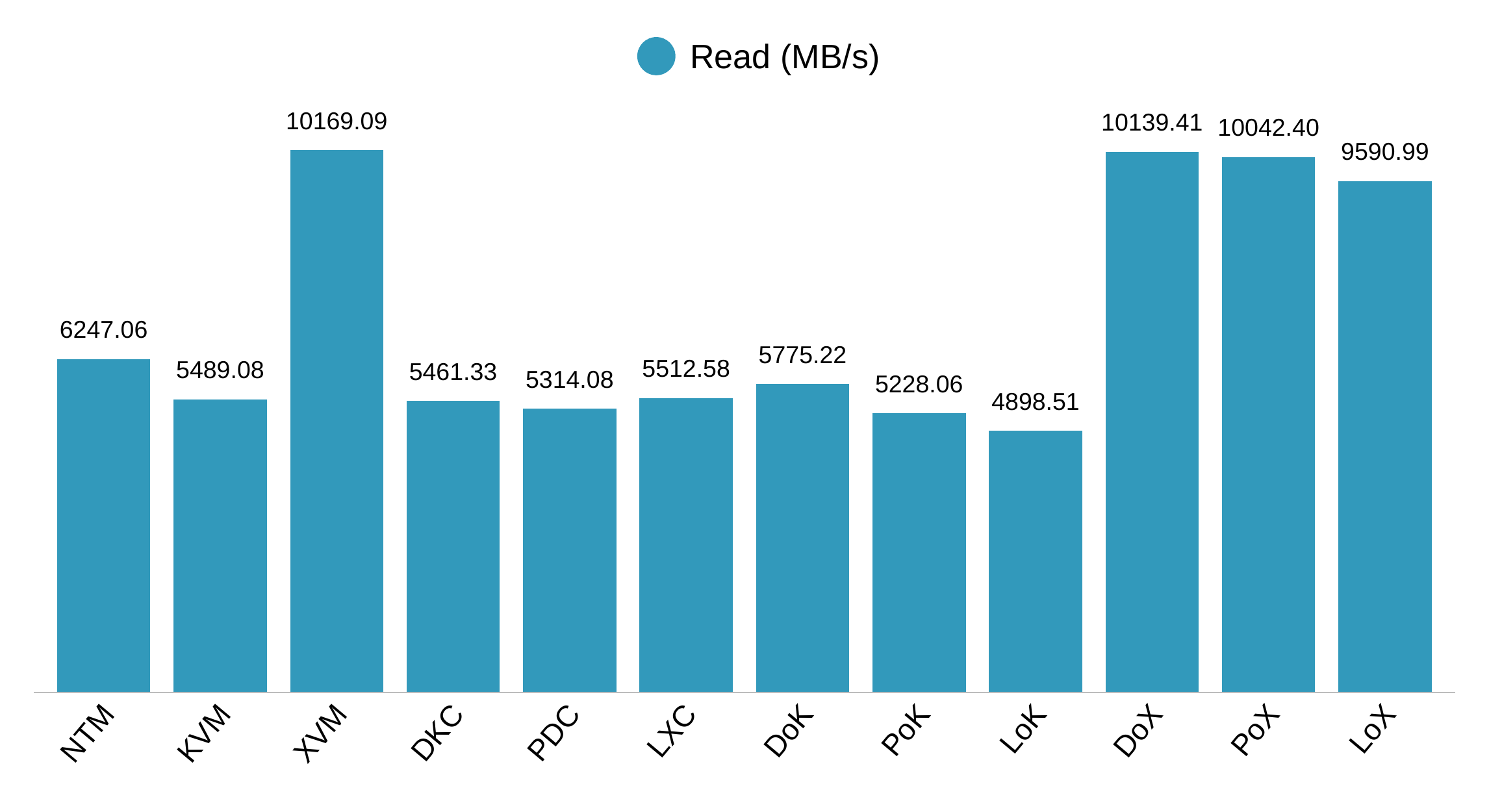}
	\end{minipage}%
	\begin{minipage}[c]{.3\linewidth}
		\captionsetup{justification=raggedright, singlelinecheck=false}
		\makebox[0pt][l]{\parbox{\linewidth}{%
				\caption{IOzone test results: read operations}\label{fig4}
		}}
	\end{minipage}
\end{figure}
In contrast to the read operations on the disk, the Podman container does not deliver desirable performance for the write operations and performs worst among all configurations. There could be several reasons for this: 
\textit{(i) storage drivers}:
Podman and Docker use different storage drivers; Docker uses \textit{aufs}, \textit{overlay}, or \textit{overlay2}, %(depending on the host system), 
while Podman uses \textit{overlayfs}; the differences in storage drivers could affect I/O performance depending on the specific workload.
\textit{(ii) security and isolation}: Podman emphasizes running containers as non-root users to improve security. Depending on the usage scenario and configuration, this additional layer of security may affect performance compared to Docker's behavior.

\begin{figure}[h]
	\centering
	\begin{minipage}[c]{.3\linewidth}
		\captionsetup{justification=raggedright, singlelinecheck=false}
		\makebox[0pt][l]{\parbox{\linewidth}{%
				\caption{IOzone test results: write operations}\label{fig5}
		}}
	\end{minipage}%
	\begin{minipage}[c]{.7\linewidth}
		\includegraphics[width=\linewidth]{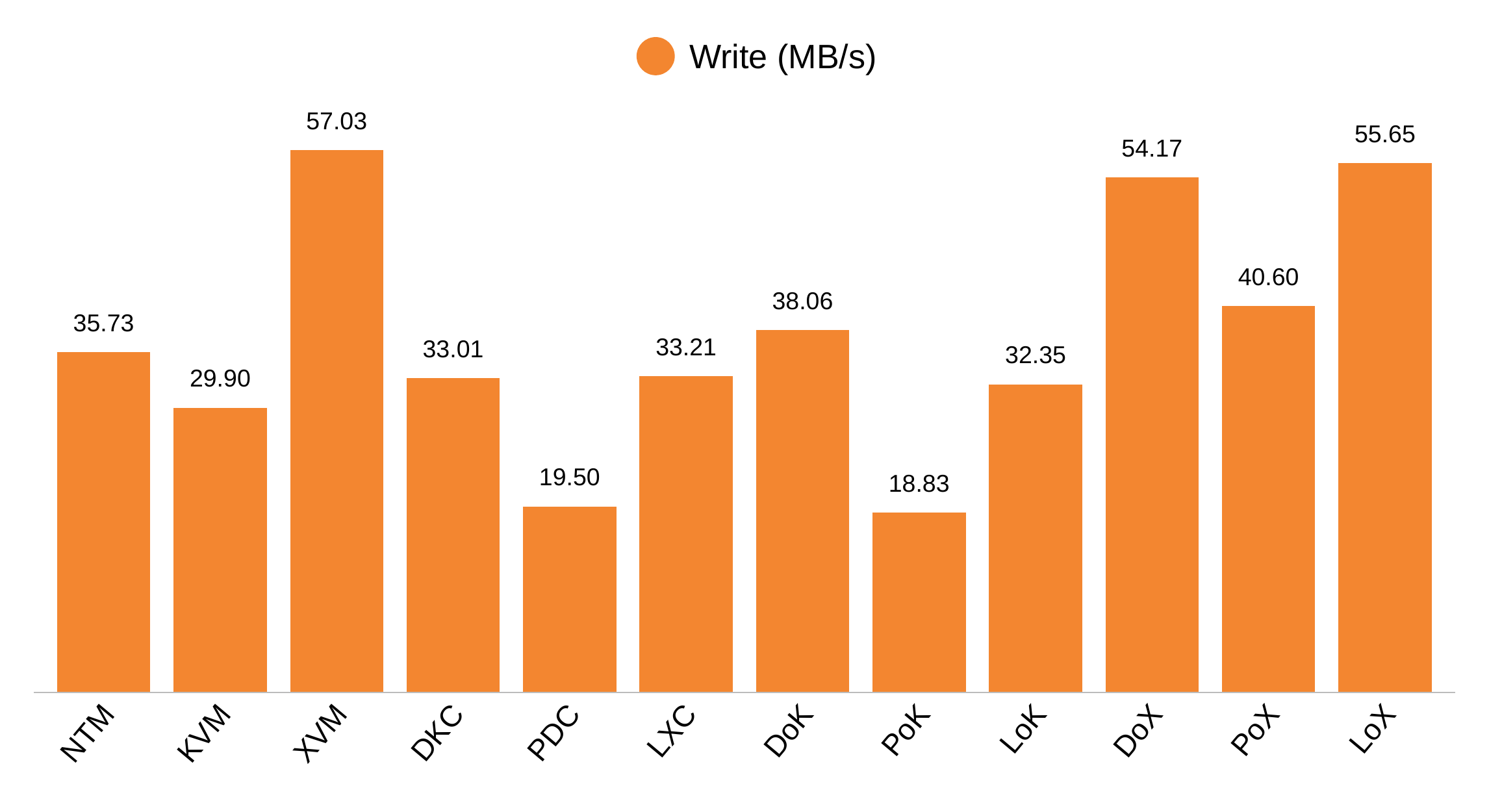}
	\end{minipage}
\end{figure}

\subsection{4.4. Network}\label{sub4sec4}
Network performance is evaluated with the Netperf benchmark. This benchmark performs tests with TCP stream, TCP request-response and UDP request-response experiments to measure and analyze network performance from different perspectives. Figure \textcolor{blue}{\ref{fig6}} illustrates the bandwidth throughput of the TCP protocol. When evaluating the TCP stream performance, almost all configurations, with the exception of the Podman container scenarios, perform almost similarly. In particular, the Podman scenarios, including the PDC itself as well as the PoK and PoX scenarios, have the lowest performance because they cannot create network interfaces on the host. 
\begin{figure}[h]
	\centering
	\begin{minipage}[c]{.7\linewidth}
		\includegraphics[width=\linewidth]{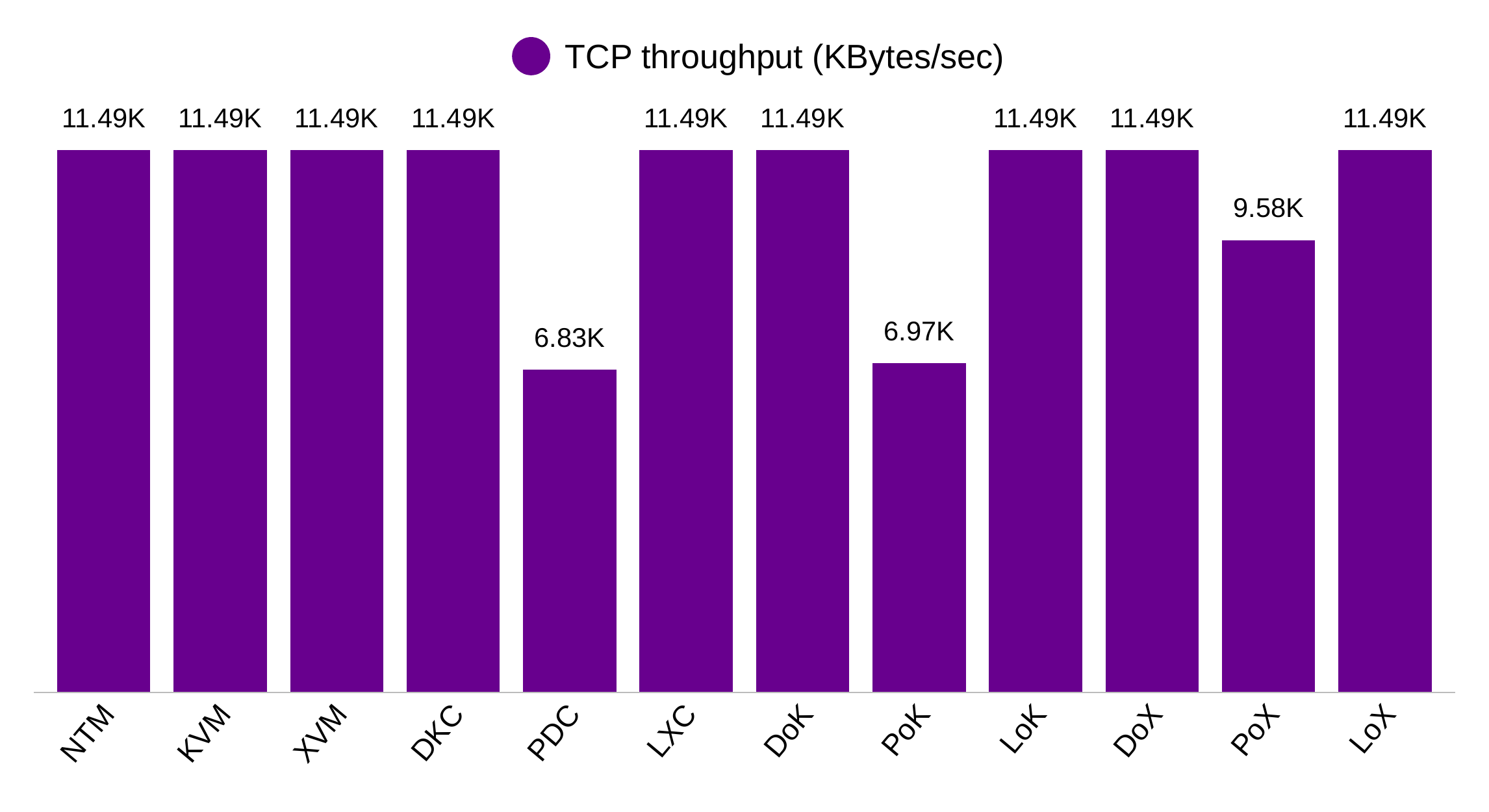}
	\end{minipage}%
	\begin{minipage}[c]{.3\linewidth}
		\captionsetup{justification=raggedright, singlelinecheck=false}
		\makebox[0pt][l]{\parbox{\linewidth}{%
				\caption{Netperf test results: TCP Stream}\label{fig6}
		}}
	\end{minipage}
\end{figure}
Containers that do not require system-level access, such as Podman, typically use the default network mode called \textit{slirp4netns}. However, it is important to note that \textit{slirp4netns} may lack some networking features, which can affect performance when benchmarking. 

Figure \textcolor{blue}{\ref{fig7}} represents the request-response rates for TCP and UDP protocols. In the tests for receiving responses, for both TCP and UDP protocols, the native machine and the Docker container perform best, followed by the Podman container. The  KVM virtual machine and the LoK perform similarly well. Interestingly, despite using the PVHVM mode, the XVM and Xen-based containers does not perform well. This indicates that Xen still has a performance overhead even with optimizations in some network operations.

\begin{figure}[h]
	\centering
	\begin{minipage}[c]{.3\linewidth}
		\captionsetup{justification=raggedright, singlelinecheck=false}
		\makebox[0pt][l]{\parbox{\linewidth}{%
				\caption{Netperf test results: TCP and UDP Request Response}\label{fig7}
		}}
	\end{minipage}%
	\begin{minipage}[c]{.7\linewidth}
		\includegraphics[width=\linewidth]{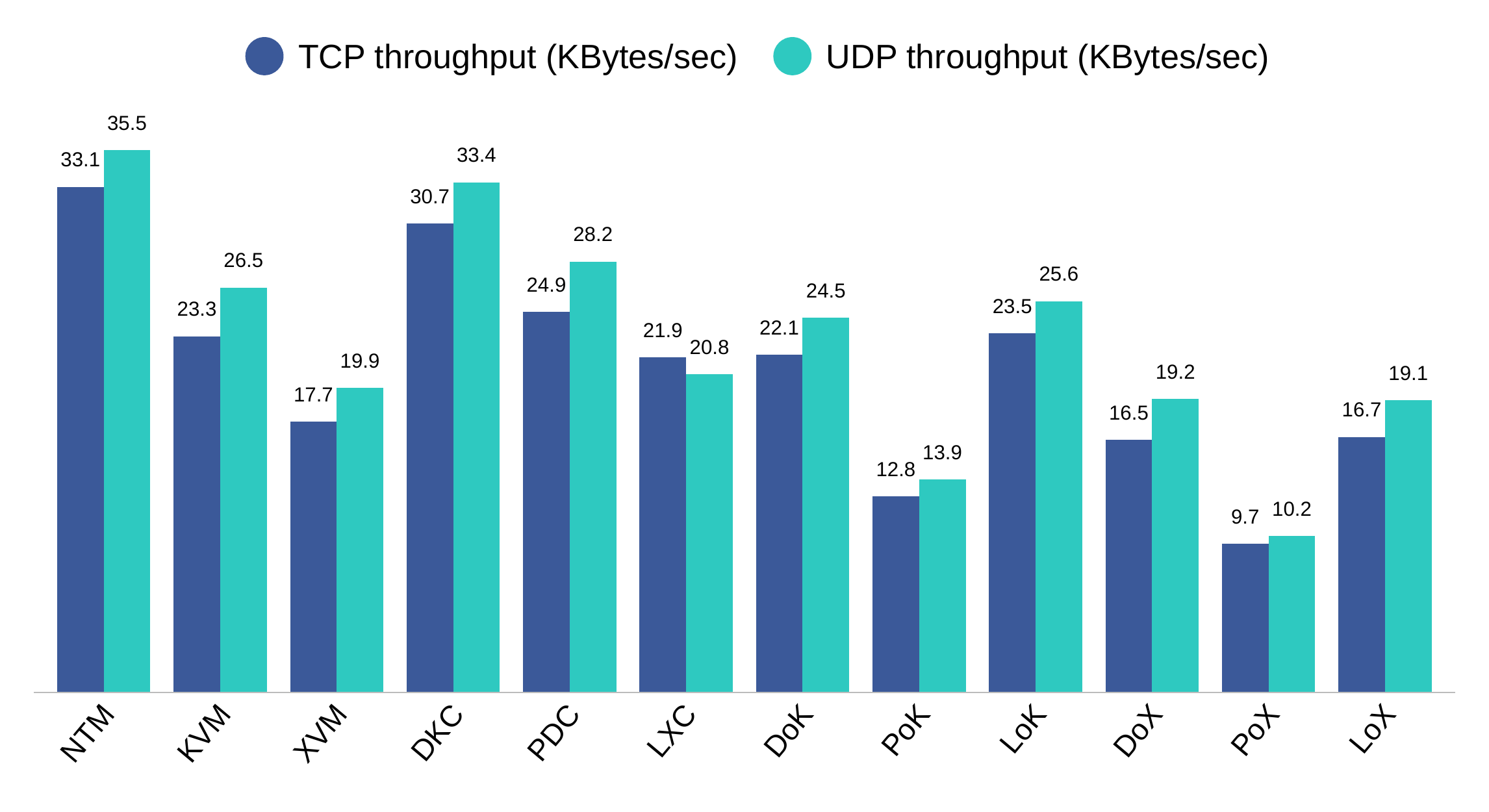}
	\end{minipage}
\end{figure}

\subsection{4.5. MySQL Database}\label{sub4sec5}

This section deals with the performance evaluation of a database management software under different configurations. In this study, MySQL, an open source database management system, is used for the experimental investigation.
Figure \textcolor{blue}{\ref{fig8}} illustrates the result of comparing the scenarios in terms of transactions per second. As can be seen in this figure, Xen-based VM and Xen-based containers perform better than the other scenarios.
Considering the close relationship between MySQL functionality and disk operations, the better performance of Xen-based scenarios compared to other scenarios can be justified. For single-threaded scenarios, the Xen virtual machine shows significantly better performance first, followed by containers on Xen (DoX, LoX, and PoX) and then other configurations that show relatively similar performance. The native machine, containers (LXC, Docker, Podman) and finally the configurations with KVM (KVM itself, PoK and LoK) follow with significant differences. As the number of parallel threads increases, the performance of Xen-based configurations increases exponentially.
In contrast, other configurations show a moderate linear growth in the number of transactions. An interesting aspect of Xen-based configurations is that up to fifty threads, there is a rapid increase in transactions per second, but beyond fifty threads and up to one hundred, the increase in transactions becomes negligible.

\begin{figure}[h]
	\centering
	\includegraphics[width=0.9\textwidth]{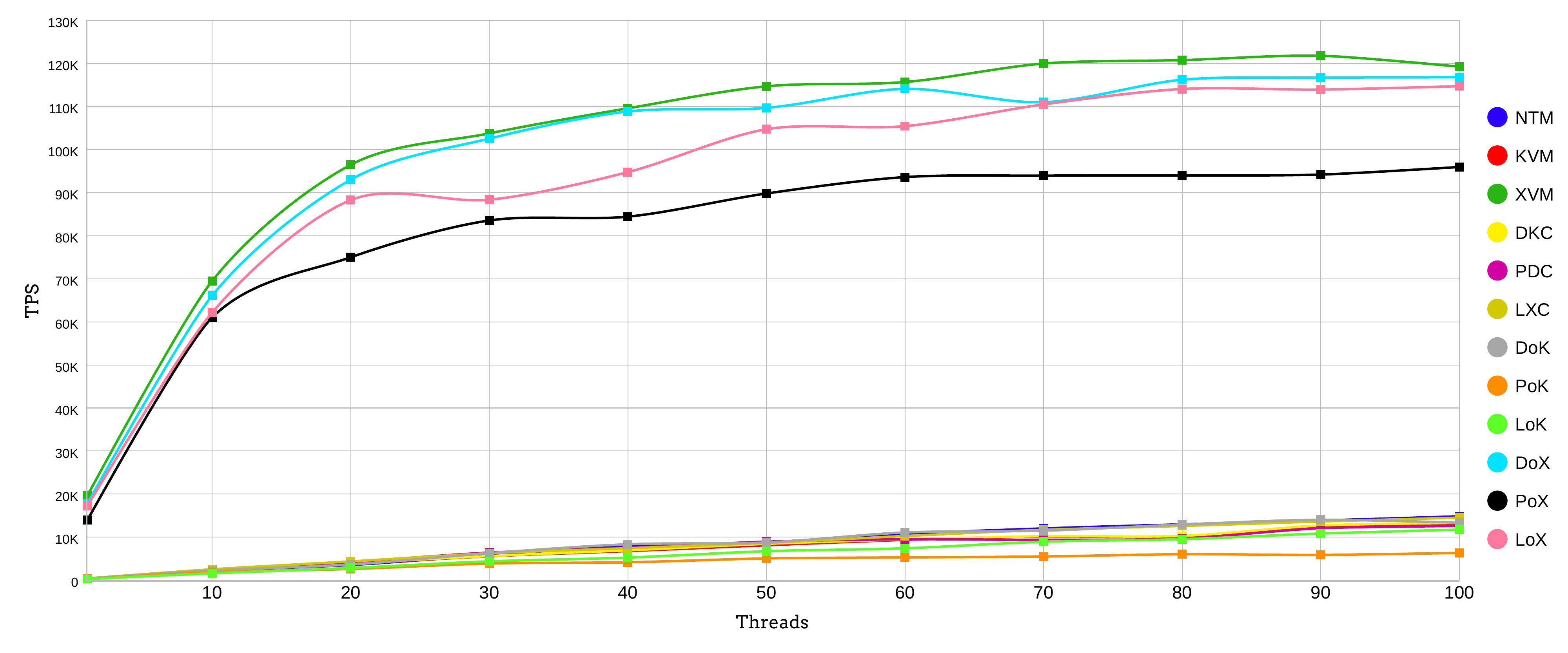}
	\captionsetup{justification=centering}
	\caption{Sysbench test results}\label{fig8}
\end{figure}

\subsection{4.6. Discussion}\label{sub4sec6}
Based on the top three rankings for each resource and benchmark, Figure \textcolor{blue}{\ref{fig9}} provides a general overview of the performance evaluation results for different scenarios.
As expected, the native machine delivers the best performance in CPU, memory and network resources, closely followed by the containers, especially Podman and Docker. 
Among configurations involving containers running over virtual machines, the combination of Docker or LXC with KVM-based VMs emerges as the optimal choice for CPU performance. In terms of memory utilization, the combination of LXC or Podman with Xen-based VMs ranks highest. Additionally, for network performance, the scenarios featuring LXC or Docker with KVM-based VMs demonstrate superior performance.

In the hard disk category, on the other hand, container configurations running on Xen, especially DoX and LoX, show the best results after the Xen-based VM. In general, the DoX configuration proves to be the best option for mixed read and write use cases. 
There are many factors that lead to a well-performing disk in Xen-based scenarios: 
\textit{(i) paravirtualization for disk}: Xen employs paravirtualization for disk operations where the guest OSs are aware of running in a virtual environment. This allows Xen to offload certain tasks to the hypervisor and reduce overhead compared to full virtualization solutions such as KVM. 
\textit{(ii) hypervisor efficiency}:  Xen is designed to be a lightweight and efficient hypervisor. It aims to achieve better performance by minimizing virtualization overhead so that VMs can achieve near-native performance in many cases. 
\textit{(iii) IO optimization}: Xen may have specific optimizations for IO operations that could be beneficial for the IOzone benchmark. These optimizations could include efficient management of disk I/O requests and better use of hardware resources.
In summary, the Xen-based VM has excellent disk performance for both read and write operations, largely due to the PVHVM mode and the overall design of Xen as an efficient hypervisor.

\begin{figure}[h]
	\centering
	\includegraphics[width=0.8\textwidth]{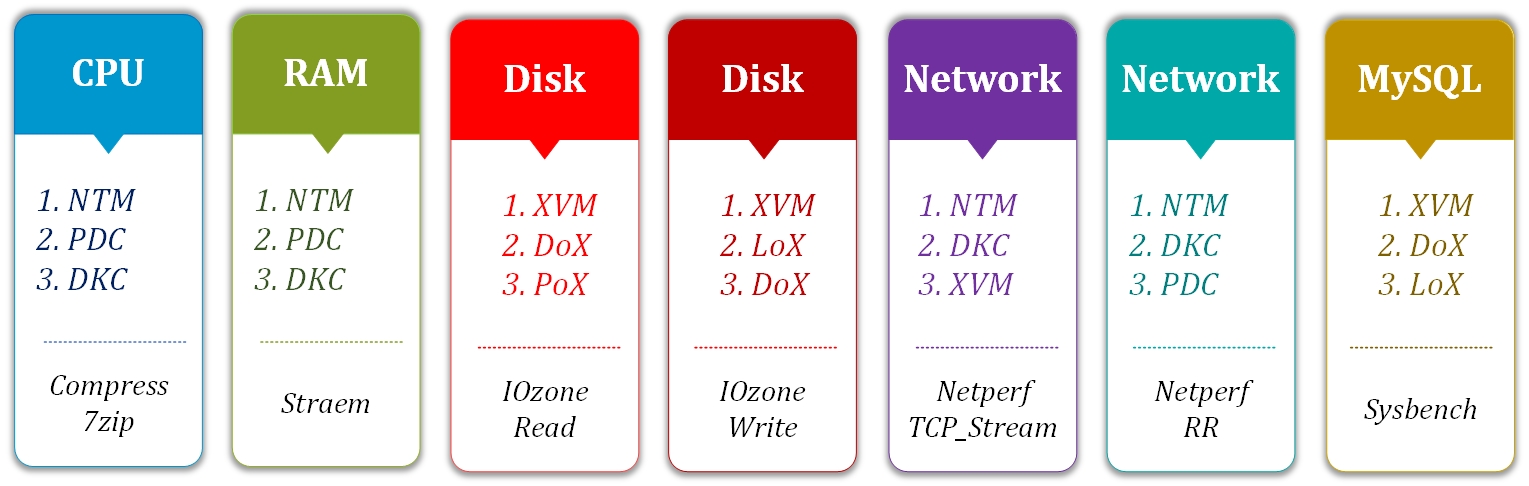}
	\captionsetup{justification=centering}
	\caption{Top-ranked scenarios based on different resources and benchmarks}\label{fig9}
\end{figure}

%%%%%%%%%%%%%%%%%%%%%%%%%%%%%%%%%%%%%%%%%%%%%%%%%%%%%
%      Conclusion
%%%%%%%%%%%%%%%%%%%%%%%%%%%%%%%%%%%%%%%%%%%%%%%%%%%%%

\section{Conclusion}\label{sec5}
In this study, we evaluated the performance of two popular and widely used virtualization technologies: virtual machines and containers. We assessed their performance separately in terms of four main resources: CPU, memory, disk, and network utilization. Various experiments were conducted under different conditions to comprehensively evaluate their performance. For this purpose, twelve scenarios were defined and examined in a real test environment. The real-world evaluations showed that containers cause less overhead and generally provide acceptable performance. In particular, Podman containers  showed the best performance in CPU and memory, while Docker containers stood out in network performance. 
Among the configurations with containers running on top of VMs, 
Docker on KVM-based VM, LXC on Xen-based VM, Docker on Xen-based VM, and LXC on KVM-based VM proved to be the optimal choice in terms of CPU, memory, disk, and network performance, respectively. 
As a real-world application, MySQL was also deployed in the mentioned scenarios, showing that the Xen-based VM and the Xen-based containers outperformed the other scenarios. 

The insights gained from this study provide guidance to data center architects and cloud engineers in selecting the most appropriate configurations for different conditions and use cases. Researchers can gain a better understanding of the strengths and weaknesses of various configurations, and cloud service providers can make informed choices based on their specific needs. The findings of this study will serve as a foundation for future work. One notable avenue for future research could be to investigate the high availability of container scenarios on virtual machines in Kubernetes clusters. Furthermore, investigating the container runtime performance of applications deployed in such configurations could be an interesting topic for further research in Kubernetes clusters.

\bibliography{wileyNJD-AMA}
\end{document}